\documentclass[twocolumn,aps,prc,showpacs,preprintnumbers,amsmath,amssymb,10pt]{revtex4-1}

\usepackage{graphicx}
\usepackage{dcolumn}
\usepackage{setspace}
\usepackage{bm}
\usepackage{multirow}
\usepackage[dvipsnames]{xcolor}

\begin{document}
\title{Obtaining high resolution excitation functions with an active thick-target approach and validating them with mirror nuclei}

\author{S. Hudan}
\author{J.~E. Johnstone}
\author{Rohit Kumar}
\author{R.~T. deSouza}
\email{desouza@indiana.edu}
\affiliation{%
Department of Chemistry and Center for Exploration of Energy and Matter, Indiana University\\
2401 Milo B. Sampson Lane, Bloomington, Indiana 47408, USA}%

\author{J. Allen}
\author{D.~W. Bardayan}
\author{D. Blankstein}
\author{C. Boomershine}
\author{S. Carmichael}
\author{A. Clark}
\author{S. Coil}
\author{S.~L. Henderson}
\author{P.~D. O'Malley}
\author{W.~W. von Seeger}

\affiliation{Department of Physics, University of Notre Dame, Notre Dame, Indiana 46556, US} %

\date{\today}

\begin{abstract}
 Measurement of fusion excitation functions for stable nuclei has largely been restricted to nuclei with significant natural abundance.  Typically, to investigate neighboring nuclei with low natural abundance has required obtaining isotopically enriched material. This restriction often limits the ability to perform such measurements. We report the measurement of a high quality fusion excitation function for a $^{17}$O beam produced from unenriched material  with 0.038\% natural abundance. The measurement is enabled by using an active thick-target approach and the accuracy of the result is validated using its mirror nucleus $^{17}$F and resonances. The result provides important information about the average fusion cross-section for the oxygen isotopic chain as a function of neutron excess.
   
\end{abstract}


\maketitle

\section{Introduction}
A new generation of radioactive beam facilities allows exploration of nuclear structure and reactions well away from $\beta$-stability \cite{FRIB, RIKEN, GANIL}. Of particular interest is the investigation of nuclear properties for extremely neutron-rich nuclei where novel low-lying collective neutron modes may exist \cite{Takigawa91, Hussein95}. Existence of these collective modes could influence nuclear reactions such as fusion (e.g. enhancing the fusion cross-section). However, beams of the most neutron-rich nuclei are typically available only at low-intensity, limiting the utilization of conventional thin-target experiments. In addition to radioactive beams, low-abundance stable beams also present a challenge necessitating the use of enriched source material which is often difficult and expensive to obtain. Fusion measurements for these low-abundance isotopes are important as they allow systematic examination of trends with increasing neutron-richness along isotopic chains enabling the investigation of pairing effects.

One solution to the challenge presented by low-intensity beams is the use of thick-target techniques. Thick targets allow measurements of cross-sections with low-intensity beams. Historically, such measurements have been restricted to cases where penetrating reaction products (e.g. $\gamma$ rays) exit the target. In recent years active thick-target approaches have been developed that enable the measurement of low-energy fusion excitation functions through direct detection of the heavily-ionizing reaction products \cite{Carnelli14,Johnstone21,Asher21}. Of particular interest is the ability to accurately characterize the fusion excitation function using these techniques and compare the result with conventional thin-target measurements to ensure the validity of the active thick-target approach.

\section{MuSIC@Indiana}
A MuSIC detector is a simple, active thick-target detector. It consists of a transverse-field, Frisch-gridded ionization chamber with the anode subdivided into strips transverse to the beam direction. The signal from each anode segment is  independently read-out allowing the energy deposit of an ionizing particle to be sampled along the beam direction. As it traverses the detector, the incident beam loses energy in the gas at a rate characterized by its specific ionization. If a fusion event occurs, amalgamation of the projectile and target nuclei results in a compound nucleus (CN) with larger atomic and mass number than the incident beam. At modest excitation the CN de-excites to form an evaporation residue (ER) which still manifests an increased atomic and mass number as compared to the beam. 
Due principally to its greater atomic number, the specific ionization of the ER is larger than that of the beam. This increase provides the ability to distinguish the presence of an ER, which signals the occurrence of a fusion event.
By determining the position at which fusion occurs within the active volume, multiple energies on the fusion excitation are simultaneously measured for a single incident energy. This approach thus provides a highly efficient means of measuring the total fusion cross-section. As the same detector is used to count both beam and evaporation residues the method is also self-normalizing.

\begin{figure}[h]
\begin{center}
\includegraphics[scale=0.500]{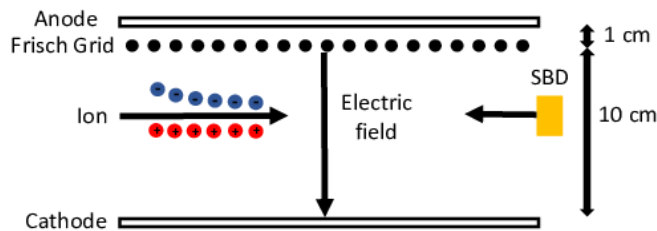}
\caption{
Schematic side view of MuSIC@Indiana. Insertion of the surface barrier silicon detector (SBD) from downstream 
into the active volume is also indicated. Taken from \cite{Johnstone21}.
}
\label{fig:MUSICDesign}
\end{center}
\end{figure}

The overall design of MuSIC@Indiana is similar to other 
MuSIC detectors presently in use \cite{Carnelli14, Asher21,Blankstein23} with some key differences. The active volume of MuSIC@Indiana is formed by six printed circuit boards (PCBs) which together constitute a rectangular box. The top and bottom of the box, with an active area of 25.97 cm x 23.00 cm, serve as the anode and cathode respectively \cite{Johnstone21}. Between the anode and cathode is a wire plane (50 $\mu$m diameter Au-W wires on a 1 mm pitch) that acts as a Frisch grid.
A side view of MuSIC@Indiana indicating the anode-to-Frisch grid and
Frisch grid-to-cathode spacings is presented in Fig. \ref{fig:MUSICDesign}. To provide a short collection time of the primary ionization produced by an incident ion, the detector is operated at
a reduced electric field of $\sim$0.7 kV/cm/atm between the cathode and the Frisch grid.
This field yields an electron drift velocity of $\sim$10 cm/$\mu$s in
CH$_{4}$ \cite{Foreman81}.
A significantly higher reduced electric field between the Frisch 
grid and the anode  ($\sim$1.4 kV/cm/atm)
minimizes termination of electrons on the Frisch grid.
Field shaping at the edges of the detector is accomplished by enclosing the active volume with four PCBs each having 1.613 mm strips with a center-to-center pitch of 3.226 mm.
 A 30 mm diameter hole in the upstream and downstream PCB allows the beam to enter and exit the active volume of the detector.
The hole in the downstream PCB also enables the precise 
insertion of a small silicon surface barrier detector (SBD) using a 
linear-translation system. This ability to insert a SBD precisely into the active volume, is critical in the calibration and operation of MuSIC@Indiana as it allows direct measurement of beam energy at a given location \cite{Johnstone21}.

The anode plane of MuSIC@Indiana consists of twenty anodes. Each anode segment is 1.219 cm wide with a 0.031 cm inter-strip separation between anodes.
This width for an anode segment along the beam direction was chosen to provide a sufficiently large $\Delta$E signal to yield a good signal-to-noise ratio. When the detector is operated at P = 150 Torr of CH$_4$ gas, an incident $^{17}$O ion with E$_{lab}$ = 50 MeV deposits a $\Delta$E of $\sim$1.5 MeV for an anode.

\section{Experimental Data}

Commissioning experiments with MuSIC@Indiana involved measurement of the fusion excitation function of $^{18}$O+ $^{12}$C \cite{Johnstone21,Johnstone22}. The $^{18}$O beam was produced from enriched source material and the beam intensity was reduced using a sieve to a rate of 10$^4$- 10$^5$ ions/s compatible with MuSIC@Indiana. Comparison of the results of these measurements demonstrated that the thick-target technique was in good agreement with prior thin target measurements \cite{Johnstone21, Johnstone22}. Having established the robustness of the approach we elected to examine the fusion excitation function for $^{17}$O+ $^{12}$C where data is relatively scarce. 
A large discrepancy in cross-section existed in the literature at E$_{cm}$ $\sim$ 14 MeV  from thin-target measurements by Eyal et al. \cite{Eyal76} and Hertz et al. \cite{Hertz78}. This discrepancy of approximately 200 mb  recently prompted a measurement of this reaction by Asher et al. \cite{Asher21a}.

The measurement by Asher et al., which used the MuSIC detector ENCORE \cite{Asher21b}, reported the presence of a large suppression of the cross-section at E$_{c.m.}$=14.3 MeV. This suppression was linked to the well known oscillatory structures in $^{16}$O + $^{12}$C and the occurrence of constructive and destructive interference resulting from a $^{16}$O + n
 + $^{12}$C type configuration was hypothesized \cite{Asher21a}. Asher et al. concluded that the discrepancy between prior data in the literature was resolved. In this work we re-examine  the large suppression at E$_{c.m.}$$\sim$14MeV as such an interference could be quite interesting as such oscillations can point to specific channel couplings.

The experiment was performed at the University of Notre Dame Nuclear Science Laboratory where a $^{17}$O beam with an intensity of 10$^4$ ions/s 
was produced from natural abundance of oxygen (0.038$\%$ $^{17}$O). The beam, accelerated to energies of E$_{lab}$= 55, 48, 47.5 and 47 MeV, impinged on MuSIC@Indiana. The CH$_4$ gas in MuSIC@Indiana, at a pressure of 150 torr (99.999$\%$ purity), served as both target and detection medium. 
At the incident energies measured, fusion of the incident beam with $^{12}$C nuclei produces $^{29}$Si with an excitation energy E$^*$=40-44 MeV. This compound nucleus de-excites by emission of neutrons, protons, and $\alpha$ particles to produce an evaporation residue. Direct measurement of the energy loss for calibration beams of $^{17,18}$O, $^{19}$F, $^{23}$Na, $^{24,26}$Mg, $^{27}$Al, and $^{28}$Si meant that the identification of residues did not primarily rely on energy loss calculations which typically have uncertainties of 10-15 $\%$ \cite{Carnelli15, Johnstone21}.

\begin{figure}
\includegraphics[scale=0.42]{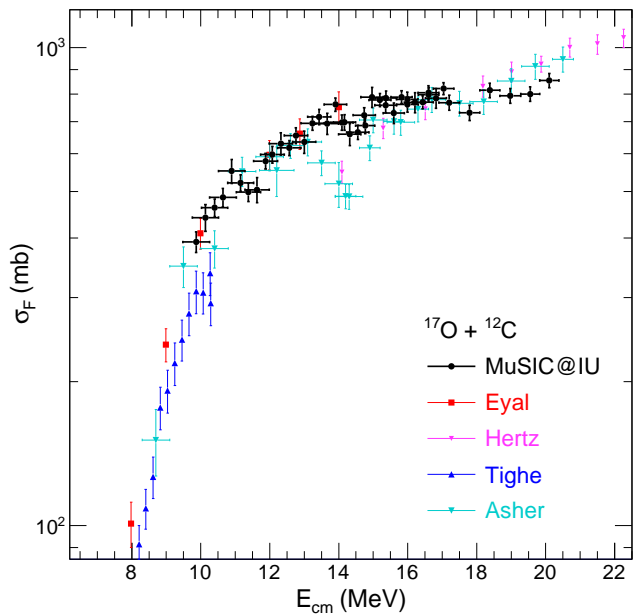}
\caption{ 
Excitation function for $^{17}$O + $^{12}$C.
}
\label{fig:17O_Data}
\end{figure}

Presented in Fig.~\ref{fig:17O_Data} is data for fusion of $^{17}$O + $^{12}$C. The present measurement spans 9.9$\leq$E$_{c.m.}$$\leq$20.1 MeV.  The emphasis in the present work of performing a high-quality measurement in the resonance regime is evident. The horizontal error bars of 300 keV, correspond to the spatial localization of the fusion event on a single anode. By utilizing small changes in the incident energy a fine mapping of the resonance regime is achieved. Vertical error bars are dominated by the statistical uncertainties. The quality of the statistics in the present measurement precluded utilizing sub-anode spatial localization \cite{Johnstone22} which would increase the cross-section uncertainty but would yield a larger number of points. 
Overall the present data are in good agreement with previously published data. Below E$_{c.m.}$=13 MeV, the cross-sections in this work largely agree, within the reported uncertainties, with those reported by Asher et al. and Eyal et al. For E$_{cm}$ $>$18 MeV the present cross-sections lie slightly below those measured by Hertz \cite{Hertz78} and Asher \cite{Asher21a}. A significant difference is observed at E$_{cm}$$\sim$14 MeV between the present data and the cross-sections recently reported by Asher et al. In the present work the dip in the cross-section
is considerably smaller in magnitude as compared to Asher et al. and
occurs at a slightly higher energy. The present results are consistent with the cross-sections previously reported by Eyal et al. \cite{Eyal76} while the Asher et al. cross-section in the resonance region is consistent with the lowest energy measurement of Hertz et al. \cite{Hertz78}. 

\begin{figure}
\includegraphics[scale=0.42]{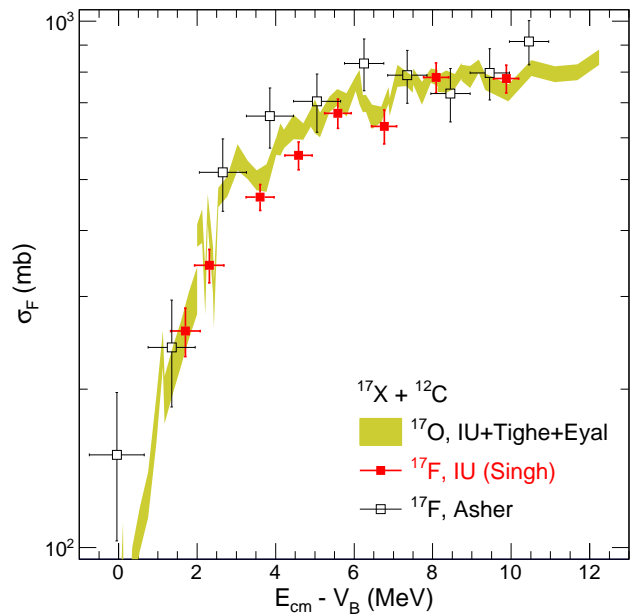}
\caption{ 
Comparison of the excitation functions for $^{17}$O + $^{12}$C with the excitation function for $^{17}$F + $^{12}$C
measured by IU \cite{Singh21a} (closed, red symbols) and Asher et al. \cite{Asher21} (open, black symbols).
}
\label{fig:Sigma_A17}
\end{figure}

In order to ascertain the accuracy of the measured cross-sections in the resonance region we compared the measured fusion excitation function with that of $^{17}$F + $^{12}$C \cite{Singh21a}. Since $^{17}$F and $^{17}$O are mirror nuclei one expects a similar behavior for the fusion excitation function when the difference in Coulomb is accounted for. Shown in Fig.~\ref{fig:Sigma_A17} are the fusion excitation functions for both $^{17}$F and $^{17}$O induced reactions. To compare the two excitation functions appropriately the cross-sections are presented relative to the Bass fusion barrier (V$_B$) \cite{DasGupta98}. For the $^{17}$O reaction the measured cross-sections of the present work together with that of Tighe et al. \cite{Tighe93} and Eyal et al. are depicted as the solid gold band. Indicated by the closed (red) symbols is the fusion excitation function (IU) for $^{17}$F + $^{12}$C \cite{Singh21a}. This measurement was a thin-target experiment which utilized an energy/time-of-flight approach to identify the ERs.  With the exception of the point at E$_{cm}$-V$_B$$\sim$5 MeV all of the $^{17}$F data lie within the gold band. In particular the resonance dip at E$_{cm}$-V$_B$$\sim$7 MeV, corresponding to the dip in the cross-section at E$_{c.m.}$$\sim$14 MeV shown in Fig.~\ref{fig:17O_Data}, is well reproduced by the $^{17}$F (IU) data. The measurement of $^{17}$F + $^{12}$C by Asher et al. \cite{Asher21} with a MuSIC detector indicated by the open (black) symbols, though systematically higher is still in reasonable agreement with the measured $^{17}$O excitation function. It should be noted that the uncertainties of the Asher et al. measurement in both cross-section and energy are significantly larger than those of the IU $^{17}$F data. This good agreement of the present $^{17}$O data with our prior  $^{17}$F measurement provides confidence in the accuracy of the present measurement of the $^{17}$O + $^{12}$C excitation function.

\begin{figure}
\includegraphics[scale=0.42]{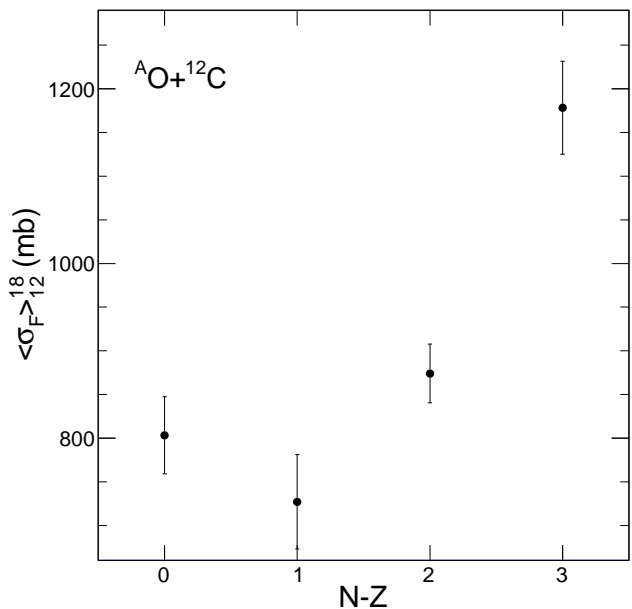}
\caption{ 
Dependence of the average fusion cross-section for $^A$O + $^{12}$C in the energy interval 12 MeV$\leq$E$_{c.m.}$$\leq$18 MeV as a function of neutron excess, (N-Z).
}
\label{fig:Avg_sigma}
\end{figure}

Having extracted an accurate excitation function for $^{17}$O + $^{12}$C, we calculated the average fusion cross-section for the interval 12 MeV$\leq$E$_{c.m.}$$\leq$18 MeV. The result is compared with the average cross-section for 
$^{16,18,19}$O + $^{12}$C  as a function of neutron excess, (N-Z) in Fig.~\ref{fig:Avg_sigma} \cite{Hudan20} The average fusion cross-section for $^{17}$O is 727 $\pm$ 54 mb, lower than that of both adjacent isotopes with paired neutrons. This reduction in average cross-section is in marked contrast to the large enhancement observed for $^{19}$O. All of the valence neutrons for (N-Z) $>$ 0 originate in the {\em{sd}} shell. This suppression of the average fusion cross-section for $^{17}$O makes the enhancement observed for $^{19}$O even more interesting as it suggests that the enhancement for $^{19}$O is not solely due to a simple understanding of pairing.

\section{Conclusions}
The fusion excitation function for $^{17}$O + $^{12}$C was successfully measured with MuSIC@Indiana using a beam from un-enriched material (0.038\% natural abundance). The resulting excitation function is in good agreement with multiple datasets in the near barrier regime. However, it
did not manifest the large suppression of the fusion cross-section at E$_{c.m.}$ $\sim$ 14 MeV observed by Asher et al. The present $^{17}$O fusion excitation function is also in good agreement with thin-target measurements of the mirror nucleus $^{17}$F when corrected for the difference in Coulomb interaction between the two systems. This agreement of the mirror nuclei further validates the accuracy of the present measurement. The extraction of the average fusion cross-section for $^{17}$O provided valuable insight into the trend of average fusion cross-section for the oxygen isotope chain. 
In summary, the present work demonstrates that active thick-target measurements can provide an effective means for accurately measuring the fusion excitation function for low-intensity beams of both radioactive or low-abundance nuclei. Such measurements are important in extracting systematic behavior along isotopic chains enabling investigation of neutron pairing effects in neutron-rich nuclei.

\begin{acknowledgments}
We acknowledge the high quality beam provided by the students and staff at the University of Notre Dame that made this experiment possible.
This work was supported by the U.S. Department of Energy Office of Science under Grant Nos. 
DE-FG02-88ER-40404 (Indiana University) and National Science Foundation PHY-2011890 (University of Notre Dame). 

\end{acknowledgments}

%

\end{document}